\documentclass[notitlepage]{article}
\usepackage{graphicx}
\usepackage{amsmath}
\usepackage{amssymb}
\usepackage{enumerate}
\usepackage{bm}
\usepackage{fancyhdr}
\usepackage[justification=raggedright,singlelinecheck=false]{caption}
\usepackage[justification=raggedright,singlelinecheck=false]{subcaption}
\usepackage{cleveref}
\usepackage{authblk}
\usepackage{verbatim}
\usepackage{makecell}
\usepackage{xcolor}
\usepackage{tikz}
\usepackage{lineno}
\usepackage{setspace}
\usepackage{glossaries}

\makenoidxglossaries

\newglossaryentry{bifurcation}
{
	name=bifurcation,
	description={A qualitative change in behaviour of a differential equation. An example of a bifurcation is the creation or disappearance of fixed points of the system in question}
}

\newglossaryentry{curl}
{
	name=curl,
	description={Measure of the local rotation in the flow}
}	

\newglossaryentry{Jacobian}
{
	name=jacobian,
	description={For an ordinary differential equation, the Jacobian is a matrix obtained by linearizing the equation at a point}
}	
\newglossaryentry{metric}
{
	name=metric,
	description={A Riemannian metric allows you to measure distances on curved space. Used here to indicate the flow down a potential may not be directly downhill}
}	
\newglossaryentry{geometry}
{
	name=geometry,
	description={Used here in the sense of topography; the formalization of a flow on a landscape}
}	
\newglossaryentry{Phenomenology}
{
	name=phenomenology,
	description={The construction of models that can directly fit observed experimental data and make predictions without necessarily manipulating or observing specific genes.}
}	

\newglossaryentry{phasespace}
{
	name=phase space,
	description={The set of variables describing the system in question.}
}		

\newglossaryentry{parameter}
{
	name = parameter,
	description = {A numerical variable which allows the model to make a range of predictions as it is varied. Biological examples include strength of a signal, rate constant in a reaction etc} 
}	

\newglossaryentry{flow}
{
	name = flow,
	description = {The flow corresponding to an ordinary differential equation is the change in the variables of the equation as a function of time usually visualized as a set of arrows signifying the direction and speed of movement in phase space}
}

\newglossaryentry{genericity}
{
	name = genericity,
	description = {Properties of the system are unchanged  when the equations are changed by a small amount in any possible way. There is an analogy with development which is robust to fluctuations}
}

\newglossaryentry{structurallystable}
{
	name = structurally stable,
	description = {A property of the flow which implies small perturbations do not change the qualitative behaviour. For example, any small change in the system should not change the number of fixed points. The connection between structural stability and development was first discussed by Thom~\cite{thom2018structural}}
}

\newglossaryentry{morsesmale}
{
	name = Morse-Smale Systems,
	description = {Dynamical systems satisfying a set of mathematical assumptions that guarantee structural stability of the flows~\cite{shub2007morse}.  We use it here in the restricted sense as a system which is structurally stable and whose long time limit (forward and backward in time) of the flows is a discrete set of fixed points
 }
}

\begin{document}


\title{A geometrical perspective on development}

\author[1,2]{Archishman Raju \thanks{Correspondence: National Centre for Biological Sciences, Kodigehalli, Bangalore 560065. Email: archishman@ncbs.res.in}}
\author[2]{Eric D. Siggia}
\affil[1]{Simons Centre for the Study of Living Machines, National Centre for Biological Sciences, Tata Institute of Fundamental Research, Bangalore 560065, India}
\affil[2]{Center for Studies in Physics and Biology, Rockefeller University, New York, NY 10065, USA}
\date{}
\maketitle
\begin{abstract}
Cell fate decisions emerge as a consequence of a complex set of gene regulatory networks. Models of these networks are known to have more parameters than data can determine. Recent work, inspired by Waddington's metaphor of a landscape, has instead tried to understand the geometry of gene regulatory networks. Here, we describe recent results on the appropriate mathematical framework for constructing these landscapes. This allows the construction of minimally parameterized models consistent with cell behavior. We review existing examples where geometrical models have been used to fit experimental data on cell fate and describe how spatial interactions between cells can be understood geometrically.  
\end{abstract}
\section{Introduction}

One of the striking facts of development is that it is \textit{canalized}. The outcome of developmental processes at different levels (cells, organs) is discrete rather than continuous. Furthermore, embryos display robustness to perturbations, and develop along well buffered paths. This is at display perhaps most strikingly in \textit{C. elegans} where every cell follows a seemingly pre-destined path which can be tracked~\cite{sulston1983embryonic}.

C. H. Waddington captured this with his metaphor of a landscape, where cell fate decisions were seen as akin to a ball rolling down a landscape~\cite{waddington2014strategy}. The landscape itself was shown as controlled by a complex network of genes. In this sense, the Waddington landscape was an early example of an emergent description: a complex network of interactions produces canalization in development. Waddington's colleague Needham explicitly talked about this in terms of emergent levels of organization, and recent historical work has uncovered the organicist philosophy that shaped their thinking~\cite{needham1943time, gilbert2000embracing}. 

As explained in Slack, the correct mathematics for understanding the Waddington landscape is dynamical systems theory~\cite{slack1991egg}. Here, levels of each gene product are represented by a differential equation, which captures interactions between genes. Developmental genetics has given us a detailed parts list and thus the modeling of such fate-decisions in terms of the underlying gene regulatory network remains popular. Nevertheless, this effort suffers from a few well-known problems. Well studied signaling pathways contain many components. Mathematical models typically write down differential equations for transduction and regulation using Hill or Michaelis Menten type of functions. These equations are commonly used but still greatly idealize the cell biology of transcription and translation. Further, they suffer from too many unknown parameters and are ill-suited to describe the data~\cite{gutenkunst2007universally}. 

Waddington's insight that the inherent complexity of gene regulatory networks leads to emergent simplicity offers the possibility of a purely phenomenological approach which tries to find the simplest set of equations which are consistent with the observed phenotypic behavior. Several people have since tried to link dynamical systems theory with Waddington's ideas. As we describe below, a general dynamical system can not be written as a gradient of a potential. One approach is to try and decompose the nonlinear dynamics of specific equations into a gradient of a potential and a remainder term which is then ignored. Early work by Huang and colleagues approximated this potential as the negative logarithm of the steady state probability distribution when noise was added to the equations~\cite{wang2010potential}. 
Later, there were other technical suggestions on how a ``quasi-potential" could be constructed through different ways to decompose a specific equation~\cite{bhattacharya2011deterministic, zhou2012quasi}. Some ignore the technical differences between a general dynamical system and one whose dynamics is determined by a potential but examine evolutionary consequences of Waddington's metaphor~\cite{jaeger2014bioattractors,saunders2018organism}. Ferrell only considered potentials in one dimension and showed how simple models of induction and lateral inhibition lead to different kinds of bifurcations~\cite{ferrell2012bistability}. Casey et al. review the link between dynamical systems theory and cell fate specification but describe Waddington's ideas as incomplete because most dynamical systems do not derive from a gradient~\cite{casey2020theory}. More recent work has tried to examine the power of Waddington's insight to understand single cell RNA-seq data~\cite{freedman2021revealing, schiebinger2021reconstructing}. Our focus in this review will be on recent work proving that typical gene networks admit a potential description provided it is supplemented with a metric~\cite{rand2021geometry}. 

From the standpoint of dynamical systems theory, there are a few qualitative possibilities for the dynamics irrespective of the underlying equations which produce the behavior. The role of mathematics is to classify those possibilities and in that process provide tools to make parsimonious fits to available data.

\section{Mathematical Concepts from Dynamical Systems Theory}

This section reviews the required concepts from dynamical systems theory and builds the analogy between developmental biology and the relevant mathematics. Definitions of technical terms used are provided in the Glossary. 

The canalization of the developmental landscape points to strong constraints in the phenotypic decision structure. Formally speaking, for a single cell, one can imagine making a list of variables which are relevant to the cell-fate decision at every step. This would correspond to a set of differential equations
\begin{equation}
\frac{d x_i}{d t} = f_i(x_1,...,x_n) 
\end{equation} 
with $i = 1,...n$. In principle, the right hand side can depend on all variables $x_i$ that have been identified to be relevant to the decision at that stage. This could include genes, external morphogens and other regulatory components enumerated by $i$ and a total of $n$ in number. These equations will produce a flow which captures the motion in this $n$ dimensional space. In a later section, we will consider how multiple cells and interactions between them can be included in the framework we describe below.

To classify the flow of a dynamical system, we need to characterize the fixed points, at which the the flow stops. There are three types of fixed points that are relevant: attractors or stable fixed points, saddle points and unstable fixed points. Biologically speaking, attractors correspond to differentiated cells at the relevant time-scales. Nearby flows all go into the attractor. Saddle-points are points where the flow is attracting in some directions and repelling in others. These correspond to undecided cells but are also the points at which the cell is most susceptible to perturbations. Finally, the flow can also have unstable fixed points which repel in all directions. It must have one such point at infinity representing boundedness of the system i.e. all genes relax to some finite value eventually. 

To classify the flow further requires that it has a discrete number of fixed points and further that it is \textit{structurally stable}. A dynamical system is said to be structurally stable if a small perturbation does not alter the qualitative behavior of the system. Biologically speaking, structural stability is to be expected from the robustness of development. Rene Thom was the first to make the link between development and structural stability~\cite{thom2018structural}. \textit{Morse Smale} systems are a class of systems which satisfy these properties~\cite{shub2007morse}. A theorem by Smale shows that it is always possible to write the flow of such systems as the gradient of a potential function with the caveat that it must be multiplied by a metric~\cite{smale1961gradient}. Thus, in general, it is always possible to write the flow as follows
\begin{equation}
\frac{d x_i}{d t} = -\sum_{j=1}^m g^{-1}_{i j} (x_k) \partial_j V(x_l)
\end{equation}
Here, $x_i$, with $i = 1,..m$ are abstract variables capturing gene activity, $V(x_l)$ is the potential, and $g_{i j}$ is the metric both of which, in principle, can be a function of all the coordinates which is labeled here with arbitrary indices $l$ and $k$. The metric $g_{i j}$ is a positive definite symmetric matrix. This theorem formalizes the idea of the Waddington landscape and shows that it is in fact possible to build a landscape for generic flows in development. The fixed points of the flows are captured in the potential. The metric captures the fact that the flow is not directly downhill i.e. it does not necessarily follow the path of steepest descent but rather may take a path which is winding. The metric also breaks the symmetry in the Jacobian of the dynamics given by a pure potential. We note here a very distinct theorem that flow can be divided into a gradient and a curl. A decomposition of the flow into this form has been used to construct a pseudo-potential which ignores the curl term~\cite{wang2010potential,zhou2012quasi}. These curl terms can be captured in the metric. 

\begin{figure}[ht]
\begin{center}
  \includegraphics[width=.9\linewidth]{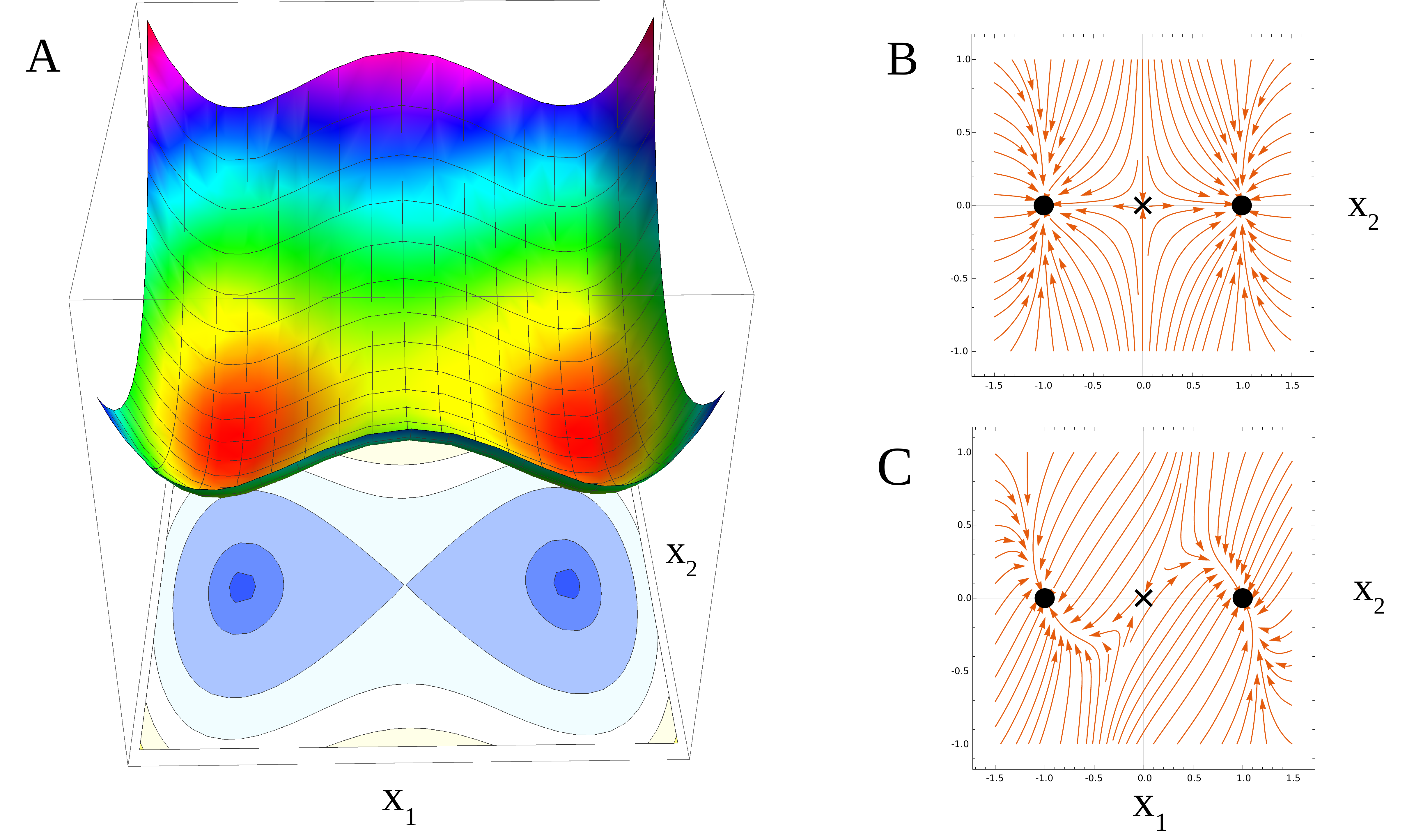}  
\caption{(A) The Waddington landscape corresponds to potential which is shown here both as a 3d plot and as a contour-plot. Both are equivalent ways to look at the potential which defines the flow. (B) The gradient of the potential defines a \textit{flow}. This flow can be visualized using streamlines. (C) It is possible for there to be a metric which can significantly alter the flow as shown here. All figures have the potential $\frac{x_2^2}{2} - \frac{x_1^2}{2} + \frac{x_1^4}{4}.$ The metric here is given by a matrix $((1 + x_2^2, 1-\epsilon), (1-\epsilon, 1+x_1^2))$ with $\epsilon$ small but positive. Here and henceforth, stable fixed points are represented by filled circles, unstable fixed points by empty circles and saddles by crosses.}
\label{potential}
\end{center}
\end{figure}

Figure~\ref{potential} gives an example in 2 dimensions with only two variables $x_1$ and $x_2$. The flow in Figure~\ref{potential}C has a non-zero curl or local rotation which implies that it is not a simple gradient flow. Nevertheless, the flow is captured by a simple bistable potential shown in Figure~\ref{potential}A, B with a metric given by $((1+x_2^2, 1-\epsilon), (1-\epsilon, 1+x_1^2))$  with $\epsilon$ small but positive. Close to the saddle point at the origin ($x, y = 0$), the metric is given by $((1, 1-\epsilon), (1-\epsilon, 1))$, and the bistable potential has Jacobian $((1, 0), (0, -1))$. The product of these two symmetric matrices produces a matrix which is not symmetric corresponding to the Jacobian of the flow shown in Figure~\ref{potential}C. Though this example is artificial, it captures the basic role of the metric and the potential. Both Figure~\ref{potential} B and C have the same set of fixed points but the flow between them is very different. Yet, both can be captured by the potential shown in Figure~\ref{potential} A with different metrics. Hence, the Waddington landscape by itself only determines the flow up to a metric.  

From an experimental point of view, it is not just the landscape itself which is interesting, but how it can be altered. Both mutations and external environmental inputs can alter the landscape. To capture this, one needs to \textit{parameterize} the landscape. The mathematics is agnostic to the biological identity of the parameters, which could be decay rates, the strength of lateral inhibition, coefficient of activation for a transcription factor, or the kinetics of the interaction between receptors and signaling proteins. Smooth changes in parameters can lead to sudden qualitative changes in the flow: a phenomena known as a bifurcation~\cite{guckenheimer2013nonlinear}. It turns out that there are only two bifurcations that one needs to consider in order to interconvert any two  Morse Smale systems with the same topology, the saddle node and the heteroclinic flip~\cite{newhouse1976there}. The saddle node is a \textit{local} bifurcation which involves the creation or destruction of two fixed points. The heteroclinic flip is a \textit{global} bifurcation which changes how the stable fixed point are connected to each other but does not create or destroy any new fixed points. From the biological point of view, the heteroclinic flip is interesting because it determines where the dynamics goes once it exits a certain stable fixed point. Some have tried to apply other bifurcations, like the pitchfork bifurcation to particular examples. However, other bifurcations rely on special symmetries. For example, the pitchfork bifurcation, when slightly perturbed breaks up into saddle node bifurcations. Thus, some strong argument is needed as to why a biological system is poised at a symmetric point.

Restricting to these two bifurcations allows the enumeration of different possibilities for the dynamics at least for situations with a small number of fixed points. In particular, for three stable fixed points and a two-dimensional phase space, it is possible to fully enumerate the simplest possible landscapes by making logical arguments on how different lines of bifurcations can meet each other~\cite{rand2021geometry}. It is worth recapitulating to see what this implies. The underlying dynamics determining the behavior of a biological system is intrinsically high dimensional. The observed phenotypic behavior is canalized and involves only a few discrete possibilities. If one has a situation where there are three possible states for the cell, then irrespective of how complicated the underlying dynamics may be, it is possible to enumerate all qualitative possibilities for the dynamics if there are two external controls. There is no mathematical restriction on what these two external controls have to be. They could be any parameter ranging from signaling factors to decay rates. Different biological systems at different stages of development can be grouped into classes which show similar behavior. Identifying the correct landscape which fits the experimental data well is a tricky process and we postpone the discussion of which experiments are most informative to a later section.

Any process that involves the creation of destruction of fixed points has to be in the parameterization of the potential. However, interestingly, it is possible to use the metric to change the connection between different fixed points. Since a heteroclinic flip, which is a global bifurcation, also acts on the trajectory between fixed points and can change the connection between them, it is possible to parameterize this bifurcation both in the potential and the metric. There is a certain redundancy in the description and it remains an open question what the best parameterization of the dynamics is and where the metric is most useful. Models with a minimal number of parameters are most useful for fitting data in a predictive manner.  

\section{Examples}

\begin{figure}[ht]
\begin{center}
  \includegraphics[width=.9\linewidth]{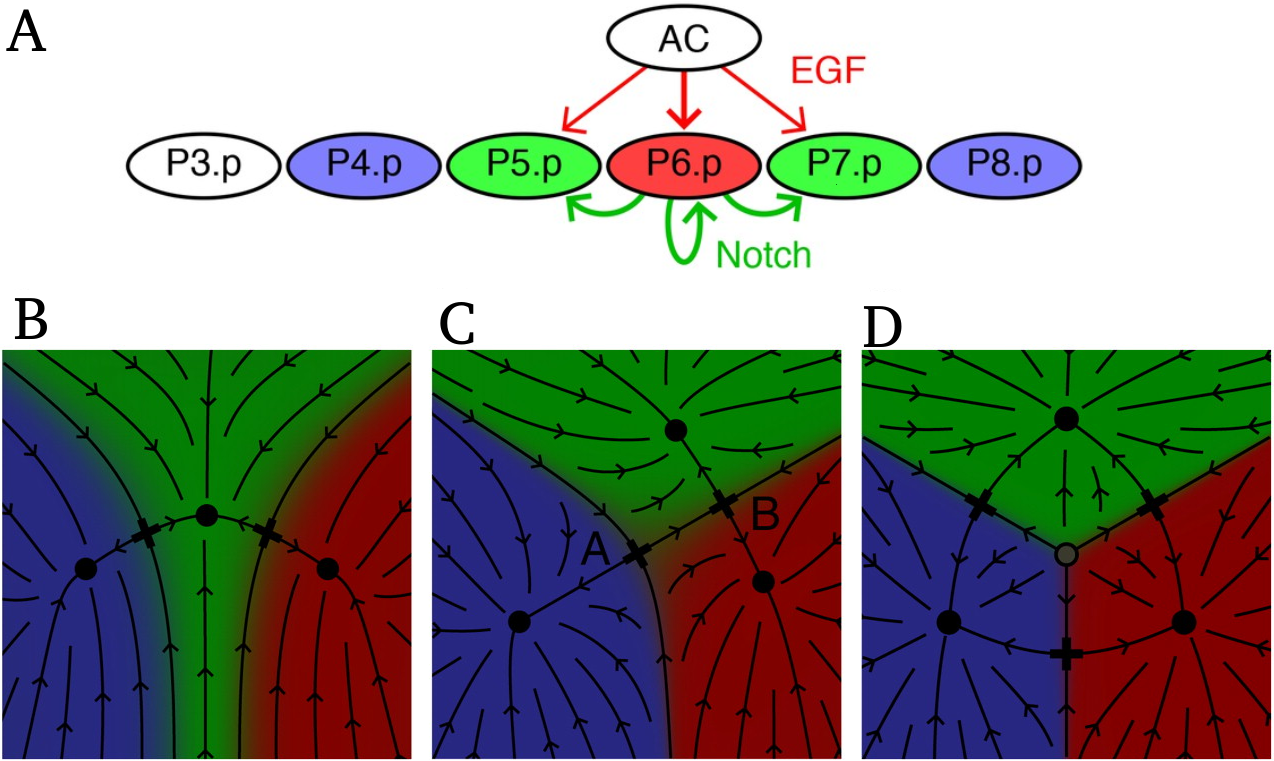}  

\caption{(A) Vulval development in C.elegans proceeds from a set of vulval precursor cells P3.p-P8.p. P3.p is special and was not considered in the model. P4.p-P8.p adopt their fate through a combination of EGF signaling from the Anchor Cell (AC) and lateral Notch signaling. The cells choose between three possible fates, $1^\circ$ or inner vulval (red), $2^\circ$ or outer vulval (green) and $3^\circ$ or non-vulval (blue). (B-D) Different possible topologies for fate decision in vulval precursor cells. The fixed points correspond to different cell fates. Note that the axes are deliberately not specified because only the fates are being modeled. In the absence of any signal the default fate is $3^\circ$ (blue). (B) The three fixed points could be in a line. EGF would push the fate towards the red region. This would mean that if the anchor cell was ablated at different times, one would obtain different proportions of green and red. To the contrary, if the anchor cell is ablated prior to a certain time, equal proportion of red and green fates are obtained ruling out this model. (C) and (D) are both consistent with the system but D is more robust and is related to the elliptic umbillic in Thom's catastrophe theory. Figure adapted from Ref.~\cite{corson2012geometry}}.
\label{vulvatop}
\end{center}
\end{figure}

The approach outlined in the previous sections is best understood by a series of concrete examples. In this section, we summarize the available biological examples which have employed the concepts developed above in specific cases. Each of these examples involve a decision between a few cell types controlled by a known signaling pathway(s). Biological data is available in the form of proportions of end point fates as the signals are perturbed. The model is thus constructed in an abstract space where the axes are arbitrary and only fates are modeled. The experimental data is used to identify the correct geometry, fit the available data and make predictions for new experiments.

\subsection{Vulval development in C. Elegans}
Corson and Siggia studied the development of the vulva in \textit{C.elegans} which follows a stereotyped development~\cite{corson2012geometry, corson2017gene, camacho2021quantifying}. A row of vulval precursor cells choose between three possible fates ($1^\circ, 2^\circ, 3^\circ$), one of them is non vulval ($3^\circ$) and the others go on to make the full vulva. In the absence of signaling, the cells all take the non-vulval fate. However, the cell fates are controlled by two signaling pathways, an inductive EGF signal which is given by an anchor cell and lateral Notch-Delta signaling. 

Corson and Siggia modeled this system using a minimal geometric model which had three possible fates and was controlled by two pathways. The first step was to determine the topology: how the three fixed points were connected to each other. As shown in Fig~\ref{vulvatop}, the three fixed points could be in a line, or involve two consecutive decisions or be symmetrically distributed with all transitions between fates possible. 

The most useful experiments in deciding between these different topologies were time-dependent anchor-cell ablation experiments. If the anchor cell is ablated early, all cells take the default fate $3^\circ$. If the anchor cell is ablated late, the wild-type phenotype is obtained. If the anchor cell was ablated at intermediate times, it led to equal proportions of the $1^\circ$ and $2^\circ$ fate. If the three fates had been in a line as shown in Figure~\ref{vulvatop}B, one would instead expect varying proportions of $1^\circ$ and $2^\circ$ depending on when the anchor cell was ablated. Corson and Siggia finally decided to model the system with the topology shown in Figure~\ref{vulvatop}D. 
They could both fit available end-point fate data on perturbations of EGF and levels of Notch ligands and further predict fate outcomes for specific timed perturbations to these signals. 

\subsection{Fate Regulation in the early Mouse Blastocyst}

\begin{figure}[ht]                                              
\begin{center}
  \includegraphics[width=.9\linewidth]{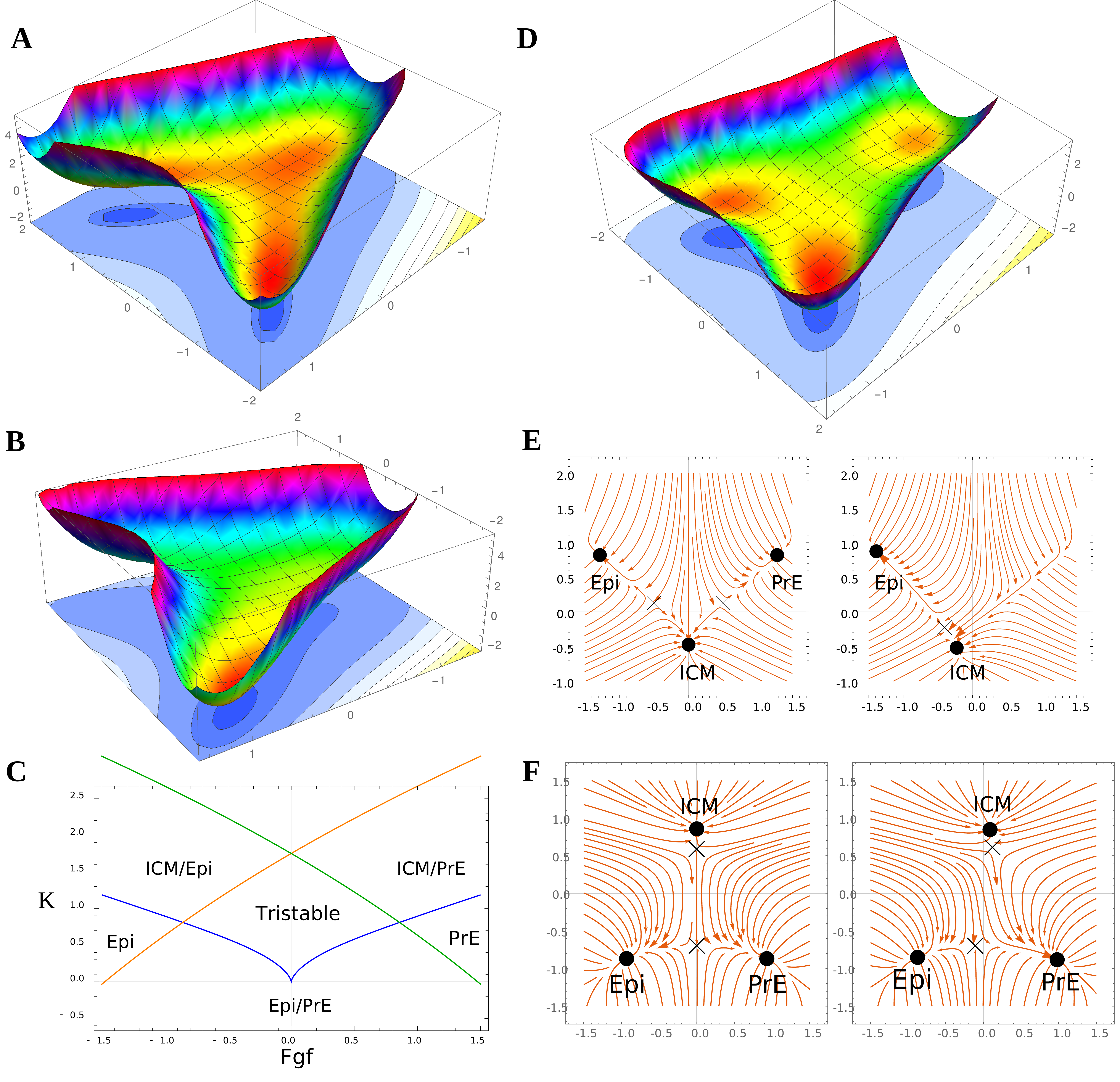}  

\caption{Two possible geometries for the decision in the blastocyst. The first is a dual cusp and the other is a heteroclinic flip. The axes for all figures except C) are arbitrary as only the fates are being modeled. A) Potential along with its contours drawn for a dual cusp. There are three minima and the center minima links the other two which have a barrier preventing direct transitions between them. B) Tilting the potential in one direction creates a \textit{bifurcation} and one of the minima disappears qualitative changing the dynamics. However, any cell exiting that minima will only go to the center minima since the barrier continues to prevent direct transitions. C) In the context of the blastocyst, the center minima corresponds to ICM and the other two minima are Epi and PrE. By considering the potential to be parameterized by Fgf and another parameter $K$ (e.g. inhibition of Nanog), one can draw a state diagram with solid lines representing different bifurcations of the minima. Negative values of Fgf indicate inhibition. As long as the ICM state exists, to go from Epi to PrE requires transitioning through the ICM state. An existing model uses this geometry~\cite{tosenberger2017multiscale}. D) The potential for the heteroclinic flip, which may be a better fit to the blastocyst dynamics. E) The flow diagram for the dual cusp. The flow is shown in two dimensions but is effectively one dimensional. Cells can transition from ICM either to Epi or PrE. Varying parameters can lead to a bifurcation where the PrE state disappears. F) The flow diagram corresponding to the heteroclinic flip. The center minima corresponds to ICM. Cells first transition out of this state before deciding between PrE and Epi. Direct transitions between PrE and Epi are allowed. Tilting the potential leads to tilting the flow lines and the dynamics becomes biased towards the PrE state. }
\label{blasto}
\end{center}
\end{figure}

The development of the blastocyst in the mouse is another example in a different context with more cells involved. The early stages of mouse development is characterized by two binary decisions, the differentiation of blastomeres into Inner Cell Mass (ICM) and Trophectoderm, and the subsequent differentiation of the ICM into Primitive Endoderm (PrE) and Epiblast (Epi) which goes on to become the embryo proper~\cite{simon2018making}. The pre-implantation mouse blastocyst is an experimentally accessible system because it can be manipulated outside of the mother~\cite{saiz2020growth}. The second of these two binary decisions is known to be controlled by the Fgf signaling pathway~\cite{yamanaka2010fgf}. Over-expression of Fgf is known to lead to all PrE fate while Fgf inhibition or Fgf receptor knockout leads to all Epi. Recent live imaging of Erk, which is downstream in the Fgf pathway, has shown that there is a considerable amount of heterogeneity in the Fgf which nevertheless translates into robust allocation of PrE and Epi fate~\cite{simon2020live, pokrass2020cell} at the embryo level. The fates are studied with the help of two markers, Nanog and Gata6. In the space of these two markers, the cell initially obtains a double positive state where both marker expressions are high, and the state then resolves itself into either high Nanog (Epi) or high Gata6 (PrE). The blastocyst literature has a discussion on the nature of the landscape involved~\cite{simon2018making}. A recent mathematical model proposed a tristable landscape which was able to qualitatively explain available experiments which modulated exogenous Fgf and/or added an Fgf inhibitor at different times~\cite{de2016cell, tosenberger2017multiscale}. This model, when studied in a two-parameter space corresponds to a particular geometry called the dual cusp with the ICM as a middle state~\cite{rand2021geometry}. The dual cusp is the confluence of two saddle node bifurcations and thus requires tuning two parameters to hit it. 
However, two possibilities exist for the landscape, the dual cusp and the heteroclinic flip as shown in Figure~\ref{blasto}. More sensitive time dependent experiments would have the potential to decisively distinguish between the two. Evidence in favor of a particular geometry could eliminate specific gene centric models which simply do not contain that geometry.

\subsection{Quantifying fate decisions in an \textit{in vitro} Embryonic Stem Cell system}

Finally, recent work on mouse embryonic stem cells (mESCs) showed how geometric models can be used to fit data with more than one decision involved~\cite{saez2022statistically}. FACS data with 5 markers was collected over 5 days. This was clustered to obtain 5 distinct cell populations as well as a set of transitioning cells, Anterior Neural (AN), Epiblast (Epi), Caudal Epiblast (CE), Posterior Neural (PN) and Mesodermal (M).  The Wnt and Fgf signaling pathways were involved in the decision and were perturbed using external signals and inhibitors at different times. The initial decision was the allocation of cells from  Epi to either AN and CE. Saez et al. argued that this decision was sudden and involved a bifurcation. This decision was thus fit using a one dimensional potential with three minima (the epiblast being in the middle). 
The next decision from CE to either PN or M was found to be much more stochastic. This decision was fit using a different landscape, the heteroclinic flip. Note this work refers to the dual cusp and the heteroclinic flip as the binary choice and the binary flip landscape respectively. Furthermore, they were able to join the two landscapes together at the CE state to thus model two separate decisions. Saez et al. were able to fit their model to a variety of experimental conditions. They then tested and validated their model through novel manipulations of the signals for which the model now had predictions. They were thus able to capture the dynamics of the decision and its dependence on the signals with a low dimensional model with a relatively small number of parameters.

\section{Cell-signaling and spatial patterns}

\begin{figure}[ht]
\begin{center}
  \includegraphics[width=.9\linewidth]{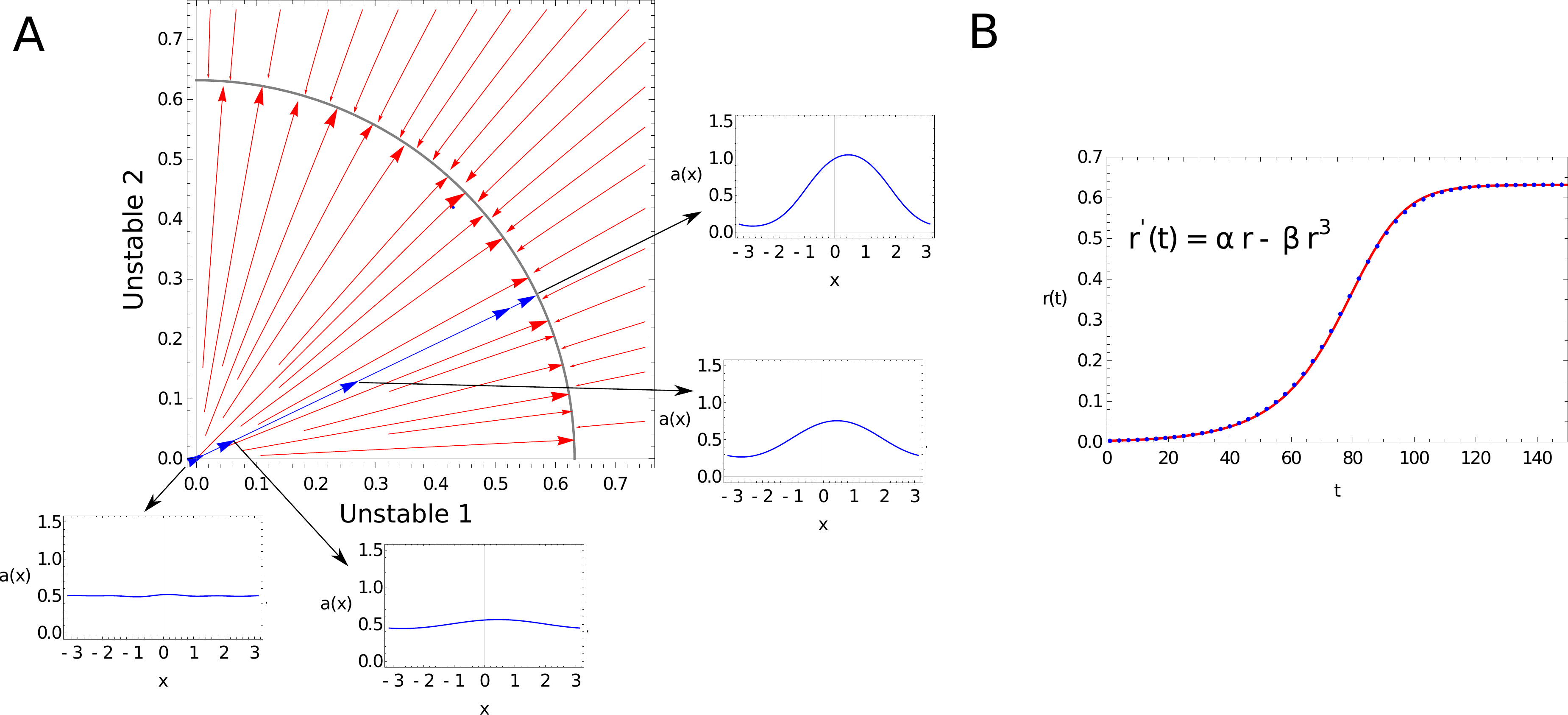}  

\caption{The geometry of a Turing system contains a few unstable directions and a large number of stable directions around the homogeneous fixed point. For an activator and inhibitor system on a bounded spatial domain, one can expand the spatial profile of the activator $a(x)$ and inhibitor $h(x)$ as a sum of trigonometric functions with increasing frequencies. Each allowed wavenumber is referred to as a Fourier mode. Most of these modes are stable, i.e. the amplitude of their coefficients decays with time on being perturbed and a few are unstable i.e. the amplitude of their coefficients grow with time. Here, an example is shown with one unstable mode and a symmetry so the mode can point in any direction on a circle. Thus there is a 2-dimensional space of unstable directions. 
(A)  A section of the unstable directions is shown with the homogeneous fixed point set to zero. Theoretically, one predicts flow away from the homogeneous fixed point to terminate at a new fixed point corresponding to the patterned state. Red arrows show streamlines obtained by fitting parameters of a polynomial potential (with a universal form) to the flow. Blue arrows show the dynamics of an activator-inhibitor system projected on to the two unstable directions at time points with uniform spacing between them. The insets show the spatial pattern of the activator $a(x)$ corresponding to these points. (B) The motion in (A) can be written in radial coordinates with increasing radius $r(t)$ and constant angle. The experimental data can be written as a {\it time independent} function of $r(t)$.
Blue points show the dynamics of this radial coordinate as a function of time for one particular trajectory marked as blue in (A). This dynamics can be written as a flow down a potential with a universal form. The red line shows the fit of the dynamics to this potential. Equations used for the fit which derive from the potential are also shown with $\alpha$ and $\beta$ parameters that are fit to. }
\label{turinggeom}
\end{center}
\end{figure}

So far the examples have only incidentally touched on the question of cell-cell communication in patterning. The landscapes mentioned above are for a single cell. Indeed, there is an ambiguity in the Waddington landscape picture itself which is sometimes used both to show the landscape of a single cell or that of a whole organism as Waddington himself did. What does adding cell-cell communication do to the landscape? This question is already implicit in the blastocyst for cells produce Fgf and it is used as a means of communication. Our unpublished analysis of recent Erk data however shows that the Fgf does not have any strong spatial correlations. It is thus possible to include the Fgf produced by all other cells as a ``mean-field" common to the entire blastocyst. This is one way to reduce the problem of many signaling cells back to one cell with the addition that the mean level (perhaps time-dependent) of Fgf enters as a parameter into the landscape and is determined self-consistently by the cell population. 

A different example was given by Corson et al. in their analysis of intermediate level Notch-Delta signaling determining the sensory organ precursors (SOP) on the dorsal thorax of the fly~\cite{corson2017self}. These SOPs are organized into rows and develop into sensory bristles on the back of the fly. They modeled the cell abstractly as comprising of two states, either SOP or epidermal. Thus, they did not explicitly model the various genes involved in the process. Cells interacting via Notch signaling organized into a row of SOP. The model was consistent with observed live imaging of reporters which first saw the emergence of stripes of proneural activity resolving into a row of SOPs. Furthermore, the model could predict the fate of Notch mutants. The dynamics was modeled with a simple sigmoidal function combined with cell-cell interactions, and can be reduced to a potential form, which reveals its essential structure. The cells compete to assume a neural precursor fate. The saddle points are the result of cells competing and neural precursor cells force non-neural behavior. The model thus starts from a homogeneous state and then transitions to the final patterned state by going through a set of saddle points~\cite{rand2021geometry}.   

Finally, the classic mechanism of self-organized pattern formation was proposed by Turing~\cite{turing1952chemical}. Older work focused on the activator inhibitor mechanism for obtaining Turing patterns, one molecule activates the formation of itself and the other inhibits the production of the activator~\cite{gierer1972theory}. While the Turing mechanism has been successful in explaining patterns, for example on fish and sea shells~\cite{kondo2010reaction}, it has been more difficult to find molecular mechanisms which correspond to the activator inhibitor framework with some exceptions~\cite{marcon2012turing}. Some have suggested looking for Turing patterns with a greater number of molecular players~\cite{sheth2012hox, raspopovic2014digit}. Others have suggested directly and abstractly modeling the interaction kernel which may incorporate different mechanisms involved in pattern formation~\cite{hiscock2015mathematically, schweisguth2019self}. Recent work in the chick embryo has revealed that mechanical forces may act as long-range inhibition leading to a Turing mechanism~\cite{caldarelli2021self}.

From the geometrical point of view, the exact nature of the molecular mechanism is not relevant for the model. The geometry of the Turing mechanism for an activator inhibitor system can be described as follows. The profile of the activator and inhibitor can be written using a Fourier series, i.e. as the sum of trigonometric functions with increasing wavenumber. The dynamics of the system can be considered in the space of the coefficients multiplying these trigonometric functions. In the absence of diffusion, there is a stable homogeneous state of both the activator and inhibitor which can be thought of as a stable fixed point. In the presence of diffusion, this homogeneous state loses its stability and both the activator and inhibitor go towards a patterned state. Geometrically speaking, the patterned state becomes a stable fixed point whereas the homogeneous state becomes a saddle point with some unstable and some stable directions. It turns out, in the presence of diffusive interactions, most of the directions corresponding to different wavenumbers remain stable: i.e. the amplitude of the coefficients decreases with time. There are only a small number of unstable directions corresponding to a few wavenumbers which is characteristic of the Turing mechanism. When seen along those unstable directions, the geometry takes on a particularly simple structure as shown in Figure~\ref{turinggeom}. It can be shown that this geometry is described by a simple potential~\cite{rand2021geometry}. Uncovering the geometry of the Turing mechanism is relevant to the description of experimental data on putative Turing systems. It shows that a simple universal potential suffices to fit the dynamics of Turing systems. It should be possible to parameterize this data using the geometry rather than refer to a molecular mechanism which may be difficult to identify. Furthermore, different molecular mechanisms may be consistent with the same geometry.

While our discussion so far has not discussed oscillations, geometrical gene-free models have also been built in somitogenesis where oscillations are inherent~\cite{franccois2012phenotypic, jutras2020geometric}. 

\section{Conclusions}

In conclusion, geometric models formalize Waddington's idea of a landscape and allow it to be used in a quantitative way to fit data. Waddington was well aware of the promise of molecular biology in understanding embryonic development. Yet, he felt than an adequate theory was required which could address the observations of experimental embryology~\cite{waddington2017concepts}. Early discussions of such an emergent theoretical description were obscured by the debate between vitalism and mechanism. Waddington and his collaborators offered a third way, one that did not appeal to vital forces but nevertheless was not reductionist~\cite{peterson2017life}.

In principle, biological data relevant to cell fate specification is high dimensional (as measured, for example by a scRNA-seq experiment). Nevertheless, typical perturbations only change the proportions of known cell types and do not create new ones. This observation can be tied to the concept of genericity and structural stability in the mathematics of dynamical systems theory. This allows the enumeration of the qualitative possibilities for the dynamics and the construction of models with a small number of parameters. In the examples we have covered, this is achieved by modeling the cell fate and not the gene expression. 

The mathematics is able to group a wide variety of developmental phenomena into similar classes. The kinds of models inspired by dynamical systems theory that we have covered in this review address the problem of having too many parameters in the model of a biological system by looking for the simplest model which is consistent with the experimental data. They are particularly well suited for time-lapse microscopy data in systems that are being perturbed by a couple of external controls. Existing molecular details, particularly available mutants, are very useful in fitting and determining the scope of the model. Furthermore, the landscape picture suggests that it is time-dependent perturbations with multiple outcomes described probabilistically that are most informative about the landscape.

\glsadd{bifurcation} \glsadd{metric} \glsadd{Phenomenology} \glsadd{Jacobian} \glsadd{curl} \glsadd{geometry} \glsadd{parameter} \glsadd{flow} \glsadd{phasespace} \glsadd{genericity} \glsadd{morsesmale} \glsadd{structurallystable}
\printnoidxglossary[nonumberlist]

\section*{Acknowledgements}
We would like to thank Raj Ladher and Arjun Guha for helpful comments on a draft of this manuscript. E.D.S. was supported by NSF Grant 2013131. A.R. acknowledges support from the Simons Foundation.

\bibliographystyle{unsrt}
\bibliography{reviewbib}

\end{document}